\documentclass[twocolumn,secnumarabic,amssymb, nobibnotes, aps, prd]{revtex4-1}

\usepackage{graphicx}
\usepackage{dcolumn}
\usepackage{bm}

\begin{document}

\title[]{Linear dust acoustic waves in inhomogeneous dusty plasmas with dust grains having power law size distribution}
\author{Gadadhar Banerjee}
\affiliation{ 
Department of Mathematics, Siksha Bhavana, Visva-Bharati University, Santiniketan-731 235, West Bengal, India.
}%
\author{Sarit Maitra}%
 \affiliation{ 
Department of Mathematics, National Institute of Technology Durgapur, Durgapur-713209, West Bengal, India.
}%
\author{Chitrita Dasgupta}%
 \affiliation{ 
Department of Mathematics, National Institute of Technology Durgapur, Durgapur-713209, West Bengal, India.
}%
\date{\today}

\begin{abstract}
The impacts of dust density inhomogeneity and dust size distribution (DSD) on linear dust acoustic (DA) wave propagation have been investigated theoretically in an inhomogeneous unmagnetized dusty plasma having power law DSD. Two different types of wave modes, viz. slow and fast mode, are found to propagate in this inhomogeneous medium. It is shown that the linear dispersion characteristics of the DA waves are substantially affected by the DSD and density inhomogeneity. Also it is found that the phase velocity increases with increasing dust density.
\end{abstract}
\maketitle

\section{Introduction}
Theoretical predictions and experimental observations of low frequency wave modes, like, dust ion acoustic (DIA)\cite{shukla2001}, dust acoustic (DA)\cite{rao1990} waves in dusty plasmas encounter rapid growth in studying different  collective waves and instabilities  in the field of homogeneous dusty plasmas\cite{nakamura1999,mendis1994,borah2016,banerjee2016,losseva2012,shukla1992}.  However, in real situations, almost everywhere in the universe, dusty plasmas are associated with various inhomogeneity in terms of number density, temparature, magnetic field inhomogenity, etc. Singh and Rao\cite{singh1998} present the linear and nonlinear study of DA wave propagation in inhomogeneous dusty plasmas by taking into account the equilibrium gradients in the plasma number density. The density inhomogeneity in plasmas can happen due to the equilibrium dust density gradient or from the equilibrium plasma density gradient. Due to density gradient, the propagation of DIA solitary waves is affected significantly in a dusty plasma having steep density profile\cite{liang2001}. The dust vortex modes in
inhomogeneous dusty plasmas  have been examined by Hasegawa and Shukla\cite{hasegawa2004}. The work of Hasegawa and Shukla\cite{hasegawa2004} is valid for unmagnetized dusty plasmas containing equilibrium electron and ion pressure gradients and the dust density inhomogeneity. Salimullah $et. al.$\cite{salimullah2004} studied the dust--lower--hybrid drift instabilities with dust charge fluctuations in an inhomogeneous dusty magnetoplasma. Using the kinetic theory, Mamun $et. al.$\cite{mamun1998ultra} explored the linear propagation of ultra-low-frequency electrostatic waves in an inhomogeneous dusty plasma. The linear dispersion characteristics of obliquely propagating  shear Alfve'n-like waves in a weakly ionized, inhomogeneous magnetized dusty  plasma are investigated by Mamun and Shukla\cite{mamun2001} where incompressible neutral fluids were considered.

Almost all of the past studies in inhomogeneous plasmas, concentrated their efforts on monosized dusty plasma for which the dust grain sizes are taken to be same\cite{singh1998,liang2001,mamun1998ultra}. In a homogeneous dusty plasma, Ma $et. al.$\cite{ma2012}  reported that the theoretical findings acquired from the system with monosized dust particles\cite{popel2005} vary from others with dust size distribution. The dust grains actually have many distinct dimensions in real situations. In space plasma, viz. cometary environments, F and G rings of Saturn, a distribution of power law\cite{meuris1997size} can describe the dust size, while a Gaussian distribution\cite{duan2007} may express it for laboratory plasmas. With distinct circumstances and environment, the size distribution may be different. Therefore, an arbitrary dust size distribution (DSD) function is essential to explore in place of monosized dust. In last few decades, different linear and nonlinear waves features have been studied in homogeneous dusty plasmas\cite{duan2003,elwakil2004,shewy2008,behery2016,labany2013,behery2015,maitra2012,maitra2014,banerjee2015}. It has been established that the wave propagation is modified due to dust mass and size variation\cite{meuris1997mass,meuris1997size}. 

Observation from different space plasmas, viz., mesospheric dusty plasma, imply the existance of both dust density  inhomogenity and DSD.  However, to the best of our knowledge, no effort has been expanded to study of linear dust acoustic waves in  an inhomogeneous dusty plasma having DSD. However, in an our recent work\cite{banerjee2017}, we have considered an inhomogeneous plasma model having DSD, but the study was confined for weakly nonlinear waves. In our present work, we have focused to study on the effects of DSD on linear dust acoustic waves in an inhomogeneous plasma having power law dust size distribution and containing Maxwellian electrons and ions and negatively charged dust particles.
dust.  Different sections of this article are organized as follows: In Sec. II, basic equations
are given. The linear theory is discussed in Sec. III. The continuous size distribution is discussed in Sec. IV and Sec. V is kept for results and discussions. Section VI presents the conclusions.
\section{Basic Equations} 

  Here we consider an inhomogeneous unmagnetized collisionless cold dusty plasma consisting of $N$ different species of dust particles with dust grain density $n_{dj}$, velocity $u_{dj}$, masses $m_{dj}$ and charges given by $Q_{dj}(=eZ_{dj})$ where $Z_{dj}$ is the number of charges residing on $j$th dust grain for $j=1,2,\dots ..N$. The steady state is inhomogeneous along $x$ direction with equilibrium density profile $N_{\alpha 0}=N_{\alpha 0}(x)$ where $\alpha =e$ for electron, $\alpha =i$ for ion and $\alpha =d_j$ for $j$th dust grain. Then the one dimensional governing equations for the dust grains are given by:
\begin{equation} \label{eq1a}
\frac{\partial n_{dj}}{\partial t}+\frac{\partial }{\partial x}\left(n_{dj}u_{dj}\right)=0
\end{equation}
\begin{equation} \label{eq1b}
\frac{\partial u_{dj}}{\partial t}+u_{dj}\frac{\partial u_{dj}}{\partial x}=\frac{Z_{dj}}{m_{dj}}\frac{\partial \phi }{\partial x}
\end{equation}
\begin{equation} \label{eq1c}
\frac{{\partial }^2\phi }{\partial x^2}=\sum^N_{j=1}{n_{dj}Z_{dj}}+n_e-n_i
\end{equation}
The electrons and ions are assumed to Boltzmann distribution given by
\begin{equation} \label{eq2a}
n_e=N_{e0}\left(x\right)\ e^{\phi \sigma s}
\end{equation}
\begin{equation} \label{eq2b}
n_i=N_{i0}\left(x\right)\ e^{-\phi s}
\end{equation}

Here, $N_{tot}(x)=\sum^N_{j=1}{n_{dj0}}(x)$ is the total number density of the dust grains, $\overline{Z_{d0}}\ =\frac{\sum^N_{j=1}{n_{dj0}}Z_{dj}}{N_{tot}}$ is average dust charge number, $\overline{m_d}=\frac{1}{N_{tot}}\sum^N_{j=1}{n_{dj0}}m_{dj}$ is average mass of dust. The different physical quantities are normalized as follows. Dust density $n_{dj}$, mass $m_{dj}$ and charge number $Z_{dj}$ of the $j$th dust grain are normalized by $N_{tot}(0)$, $\overline{Z_{d0}}$ and $\overline{m_d}$, respectively. The space coordinate $x$, time $t$, velocity $u_{dj}$ and electrostatic potential $\phi $ are normalized by Debye length ${{\lambda }_d}_{\ }={\left(\frac{T_{eff}}{4\pi e^2\overline{Z_{d0}}\ N_{tot}}\right)}^{\frac{1}{2}}$, inverse of effective dusty plasma frequency ${\omega }^{-1}_{pd}={\left(\frac{m_d}{4\pi \overline{Z_{d0}}\ N_{tot}e^2}\right)}^{\frac{1}{2}}$, the effective dust acoustic speed  $C_d={\left(\frac{\overline{Z_{d0}}\ T_{eff}}{\overline{m_d}}\right)}^{\frac{1}{2}}$ and $\frac{T_{eff}}{e}$ respectively.   Here, $T_{eff}=\frac{T_iT_e}{N_{e0}(0) T_i+N_{i0}(0) T_e}$ is defined as the effective temperature, electron and ion number densities normalized by  $\overline{Z_{d0}}\ N_{tot}(0)$, $N_{e0}(x)\ =\frac{n_{e0}(x)}{\overline{Z_{d0}}\ N_{tot}(0)\mathrm{\ }}\ $, $N_{i0}(x)=\frac{n_{i0}(x)}{\overline{Z_{d0}}\ N_{tot}(0)\mathrm{\ }}$, $n_{e0}(x)$ and $n_{i0}(x)$ are unperturbed electron and ion number density and $T_i$,$\ T_e$ being ion and electron temperatures, $s=\frac{1}{N_{i0}(0) +N_{e0}(0) \sigma }$ and $\sigma =\frac{T_i}{T_e}$. The charge neutrality condition, at equilibrium, is
\begin{equation} \label{eq3} 
N_{i0}(x)=\sum^N_{j=1}{Z_{dj0}{\ N}_{dj0}\left(x\right)}+N_{e0}(x) 
\end{equation} 

\section{LINEAR ANALYSIS}
To carry out the linear analysis for DA waves we use the method of Ducet \textit{et al.}\cite{doucet1974} where the zeroth order fields and velocities have been neglected. The system of Eqs. (\ref{eq1a})-(\ref{eq2b}) are linearized by writing the dependent variables, representing the density, speed, and electrostatic potential, as a sum of equilibrium and perturbed parts, given by, 
\begin{equation} \label{linear_nd}
n_{dj}=n_{dj0}(x)+n_{dj1}(x,t)
\end{equation}
\begin{equation} \label{linear_ud}
u_{dj}=0+u_{dj1}(x,t)
\end{equation}
\begin{equation} \label{linear_phi}
\phi=0+\phi_1(x,t)
\end{equation}
After linearization, Eqs. (\ref{eq1a})--(\ref{eq1c}) can be written as
\begin{equation} \label{linear_1}
\frac{\partial n_{dj1}}{\partial t}+\frac{\partial }{\partial x}\left(n_{dj0}u_{dj1}\right)=0
\end{equation}
\begin{equation} \label{linear_2}
\frac{\partial u_{dj1}}{\partial t}=\frac{Z_{dj}}{m_{dj}}\frac{\partial \phi_1 }{\partial x}
\end{equation}
\begin{equation} \label{linear_3}
\frac{{\partial }^2\phi_1 }{\partial x^2}=\sum^N_{j=1}{n_{dj1}Z_{dj}}+ C_{ei} \phi_1
\end{equation}
where $C_{ei}=N_{e0}(x)\sigma s+N_{i0}(x)s$.
Assuming the spatial and time dependence of all perturbed quantities vary as
\begin{equation} \label{linear_4}
A(x,t)=A_1(x) e^{-i \omega t}
\end{equation}
where $\omega$ is the angular frequency and  $A=n_{dj}$, $u_{dj}$, $\phi$, the system of Eqs. (\ref{linear_1})--(\ref{linear_3}) reduce to
\begin{equation} \label{linear_5}
-i \omega n_{dj1}+n_{dj0} \frac{\partial u_{dj1}}{\partial x}+u_{dj1} \frac{\partial n_{dj0}}{\partial x}=0
\end{equation}
\begin{equation}\label{linear_6}
-i \omega u_{dj1}=\frac{Z_{dj}}{m_{dj}} \frac{\partial \phi_1}{\partial x}
\end{equation}
\begin{equation}\label{linear_7}
\frac{{\partial }^2\phi_1 }{\partial x^2}=\sum^N_{j=1}{n_{dj1}Z_{dj}}+ C_{ei} \phi_1.
\end{equation}
Using Eqs. (\ref{linear_5}) and (\ref{linear_6}), we get,
\begin{equation} \label{linear_8}
n_{dj1}=\frac{1}{\omega^2} \left\lbrace \frac{n_{dj0} Z_{dj}}{m_{dj}} \frac{{\partial }^2\phi_1 }{\partial x^2} + \frac{Z_{dj}}{m_{dj}} \frac{\partial n_{dj0}}{\partial x} \frac{\partial \phi_1}{\partial x} \right\rbrace
\end{equation}

Now, eliminating $n_{dj1}$ from Eq. (\ref{linear_7}) and \ref{linear_8}, we obtain the following $2^{nd}$ order ordinary differential equation
\begin{equation} \label{linear_9}
\alpha \frac{{\partial }^2\phi_1 }{\partial x^2}+\beta \frac{\partial \phi_1 }{\partial x}+\gamma \phi_1=0,
\end{equation}
where 
\begin{equation} \label{linear_9a}
\alpha=\frac{1}{\omega^2} \left( \sum^N_{j=1} \frac{n_{dj0} Z^2_{dj}}{m_{dj}}-\omega^2 \right),
\end{equation}
\begin{equation} \label{linear_9b}
\beta=\frac{1}{\omega^2} \sum^N_{j=1} \frac{Z^2_{dj}}{m_{dj}} \frac{\partial n_{dj0}}{\partial x}, 
\end{equation}
and \begin{equation} \label{linear_9c}
\gamma=(\sigma s N_{e0}+s N_{i0}).
\end{equation}
\section{DUST SIZE DISTRIBUTION (DSD)}
Assuming that the radius of dust grains $a_j\ll {\lambda }_{Dd}$ the mass and charge of the dust grain can be epressed as $m_{dj}=k_ma^3_j$ where $k_m\approx \frac{4}{3\pi {\rho }_d}$ , where ${\rho }_d$ is the mass density of the dust grains. The initial charge $Q_{d0j}=k_qa^l_j$, where $k_q=4\pi {\epsilon }_0V_0$, ${\epsilon }_0$ is the vacuum permittivity and $V_0$ is the surface potential at equilibrium. Thus $Z_{d0j}=k_za^l_j$ , $k_z\approx \frac{4\pi {\epsilon }_0V_0}{e}$ and and $k_m,\ k_q,\ k_z$ are approximately constant. $l(\ge 1)$ is a constant which depends on some dust parameters and other plasma conditions. Here, we assume that the dust grains are having power law size distribution\citep{meuris1997size}. The differential from of power law size distribution is given by
\begin{equation} \label{eq24} 
n\left(r\right)dr=K r^{-p}dr 
\end{equation} 
for $r\in \left[a_{min}\ ,\ a_{max}\right]\ $and $n\left(r\right)=0$ outside the interval, where $p$ is the power law index and $n\left(r\right)dr$ is the number of the dust grains per unit volume with radii in the range from $r$ to $r+dr$. Then we have $N_{tot}=\int^{a_{max}}_{a_{min}}{n\left(r\right)\ dr}$, which gives
\begin{equation} \label{eq25} 
K=\frac{\left(1-p\right)\ N_{tot}}{a^{1-p}_{max}-a^{1-p}_{min}} 
\end{equation} 
Here, the radii of the dust grains are normalized by average dust grain radius 
\begin{eqnarray} \label{eq26a}
\overline{a}=\frac{1}{N_{tot}}\int^{a_{max}}_{a_{min}}{r\ n\left(r\right)\ dr}= \nonumber \\
\frac{1-p}{2-p}\left(\frac{a^{2-p}_{max}-a^{2-p}_{min}}{a^{1-p}_{max}-a^{1-p}_{min}}\right)=\frac{a_{min}\left(1-p\right)\left(a^{2-p}-1\right)}{\left(2-p\right)\left(a^{1-p}-1\right)}
\end{eqnarray}
\begin{equation} \label{eq26b}
a=\frac{a_{max}}{a_{min}}
\end{equation}
\noindent For inhomogeneous plasma, it is clear from the charge neutrality condition (\ref{eq3}) that the dust density also depends upon $x$. Considering the case that the DSD and inhomogeneity in dusty plasma are independent, we have joint distribution of $x$ and $r$ as
\begin{equation}
n_{dj0}\equiv n(x,r)=n_x(x) n_r(r)
\end{equation}
\noindent where $n_r(r)$ is the power law DSD function. Using the charge neutrality condition (\ref{eq3}) at equilibrium we get the normalized form of dust density function as follows,
\begin{equation} \label{eq27a}
n_x(x)=\frac{(l-p+1)(a^{1-p}-1)}{(1-p)a^l_{min}(a^{l-p+1}-1)}\left\lbrace N_{i0}(x)-N_{e0}(x)\right\rbrace
\end{equation}
and so
\begin{equation}\label{eq27b}
n(x,r)=\frac{(l-p+1)}{a^{l-p+1}_{min}(a^{l-p+1}-1)}\left\lbrace N_{i0}(x)-N_{e0}(x)\right\rbrace r^{-p}
\end{equation}
Considering continuous power law dust size distribution, the coefficients of Eq. (\ref{linear_9}) become
\begin{equation} \label{linear_9a}
\alpha=\frac{1}{\omega^2} \left\lbrace  \frac{(l-p+1)(a^{2l-p-2}-1)(N_{i0}-N_{e0})a^{l-3}_{min}}{(2l-p-2)(a^{l-p+1}-1)}-\omega^2 \right\rbrace,
\end{equation} and
\begin{equation} \label{linear_9b}
\beta=\frac{1}{\omega^2}  \left\lbrace  \frac{(l-p+1)(a^{2l-p-2}-1)a^{l-3}_{min}}{(2l-p-2)(a^{l-p+1}-1)} \frac{d}{dx} (N_{i0}-N_{e0}) \right\rbrace.
\end{equation}

\section{RESULTS AND DISCUSSIONS}
We first consider a simple case where the dust density is spatially homogeneous, $i.e.$, spatial gradients of the unperturbed  quantities are zero and so we analyzed Eq. (\ref{linear_9}) and replace $\frac{d}{dx}$ by $ik$, where $k$ is the wave number. With this assumptions,   Eq. (\ref{linear_9}) leads to
\begin{equation}
\omega^2=\frac{k^2}{k^2+\gamma} \sum^N_{j=1} \frac{n_{dj0} Z^2_{dj}}{m_{dj}}
\end{equation} which is the dispersion relation for the dust acoustic wave obtained similar to our earlier work on homogeneous dusty plasma\cite{banerjee2015}.

Next, we analyze  Eq. (\ref{linear_9}) for a more general case where spatially inhomogeneity has not been neglected. Thus, the solution of Eq. (\ref{linear_9}) can be obtain in the form of\cite{singh1998}
\begin{equation} \label{linear_10}
\phi_1(x)=\overline{\phi_1}(x) cos (kx)
\end{equation} where $\overline{\phi_1}(x)$ is the slowly varying amplitude of the wave propagating in an inhomogeneous plasma. So, by substituting Eq. (\ref{linear_10}) in (\ref{linear_9}), we get
\begin{equation} \label{linear_11a}
\alpha \frac{d^2 \overline{\phi_1}(x)}{dx^2}+\beta \frac{d \overline{\phi_1}(x)}{dx}+(\gamma-\alpha k^2)\overline{\phi_1}(x)=0
\end{equation}
and \begin{equation} \label{linear_11b}
\frac{d \overline{\phi_1}(x)}{dx}+\frac{\beta}{2 \alpha} \overline{\phi_1}(x)=0
\end{equation}
which must be satisfied simultaneously. The general solution of  Eq. (\ref{linear_11b}) can be written as 
\begin{equation} \label{linear_12}
\overline{\phi_1}(x)=C exp \left[ -\frac{1}{2} \int \left( \frac{\sum^N_{j=1} \frac{Z^2_{dj}}{m_{dj}} \frac{d n_{dj0}}{dx}}{\sum^N_{j=1}\frac{n_{dj0} Z^2_{dj}}{m_{dj}}-\omega^2} \right)dx \right]
\end{equation} where $C$ is an arbitrary constant. Considering continuous power law dust size distribution and using the charge neutrality condition at equilibrium, Eq. (\ref{linear_12}) can be written as
\begin{equation} \label{linear_13}
\overline{\phi_1}(x)=C exp \left[ -\frac{1}{2} \int  \frac{D \ d(N_{i0}-N_{e0})}{D (N_{i0}-N_{e0})-\omega^2} \right]
\end{equation}
which implies
\begin{equation} \label{linear_14}
\overline{\phi_1}(x)= \frac{C}{\sqrt{D (N_{i0}(x)-N_{e0}(x))-\omega^2 }},
\end{equation} where
\begin{equation} \label{linear_15}
D=\frac{(l-p+1)(a^{2l-p-2}-1)a^{l-3}_{min}}{(2l-p-2)(a^{l-p+1}-1)}.
\end{equation}
As Eq. (\ref{linear_11a}) and (\ref{linear_11b}) satisfy simultaneously, $\overline{\phi_1}(x)$ will be a solution of Eq. (\ref{linear_11a}). Putting the expression of $\overline{\phi_1}(x)$ from Eq. (\ref{linear_14}) in Eq. (\ref{linear_11a}), we get,
 \begin{equation} \label{linear_16}
 \left[ \beta^2+2(\alpha \beta^{\prime}-\alpha^{\prime} \beta)\right]=4(\alpha^2 k^2-\gamma \alpha) 
  \end{equation}
 which implies 
 \begin{equation} \label{linear_17}
 4 k^2 (a_1-\omega^2)^2+2(a_1-\omega^2)(a^{\prime \prime}_1-2\omega^2 \gamma)-a^{\prime 2}_1=0
 \end{equation}
 where $a_1\equiv a_1(x)=D(N_{i0}(x)-N_{e0}(x))$. Then the dust acoustic phase velocity is given by

\begin{figure*} 
(a) \includegraphics[height=6cm,width=8cm]{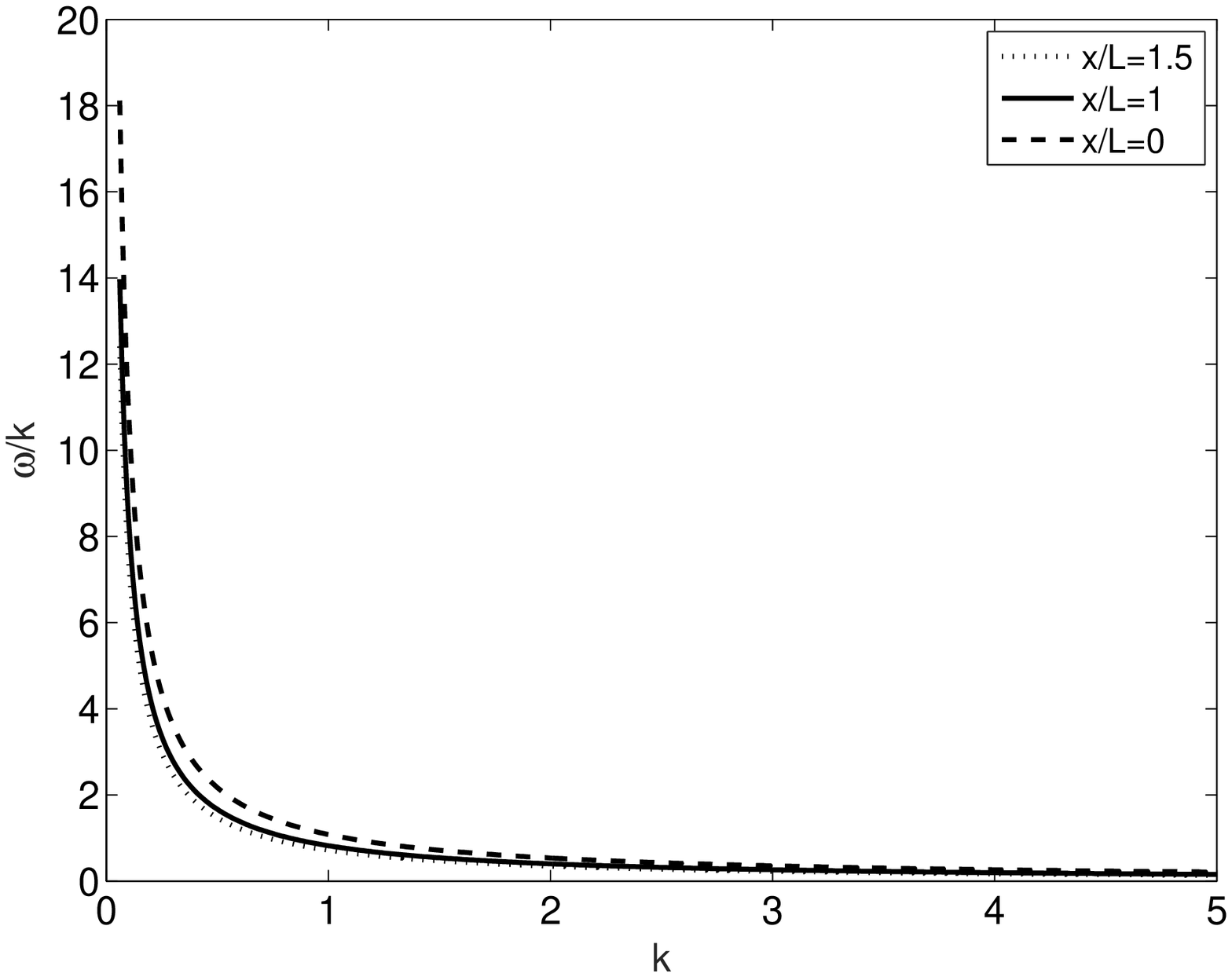}
(b) \includegraphics[height=6cm,width=8cm]{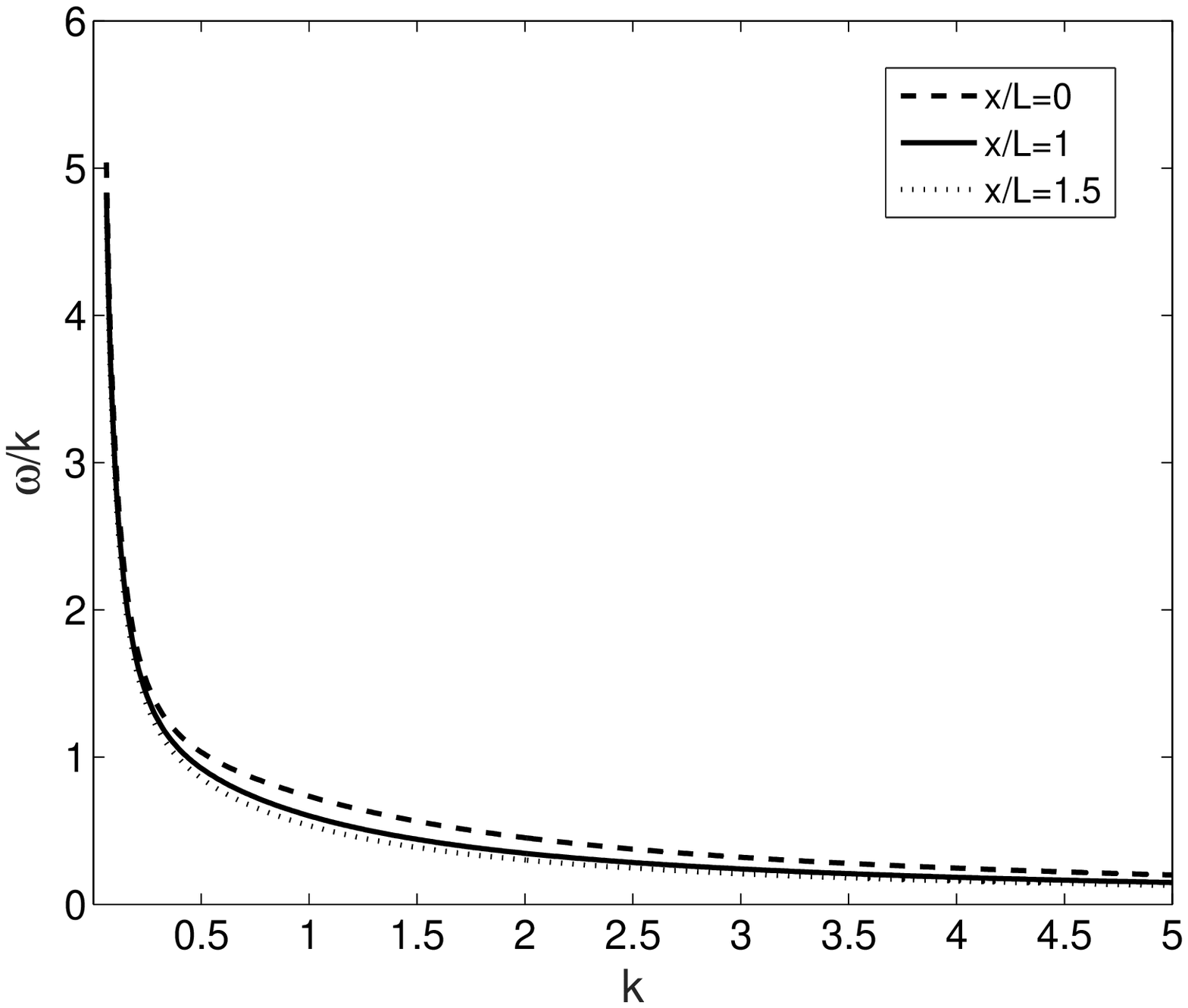}
\caption{ $\omega/k$ vs $k$ is plotted for different values of $x/L$ in (a) fast mode and (b) slow mode. Here, $l=1$, $ p=3.4$, $a=3$, $N_{e0}(0)=0.2$, $N_{i0}(0)=1.2$, $\phi_0=0$ $\sigma=0.01$, $\mu_i=0.6$, $\mu_e=0.5$.} 
\label{Fig: wk_k}
\end{figure*}

\begin{figure*} 
(a) \includegraphics[height=10cm,width=8cm]{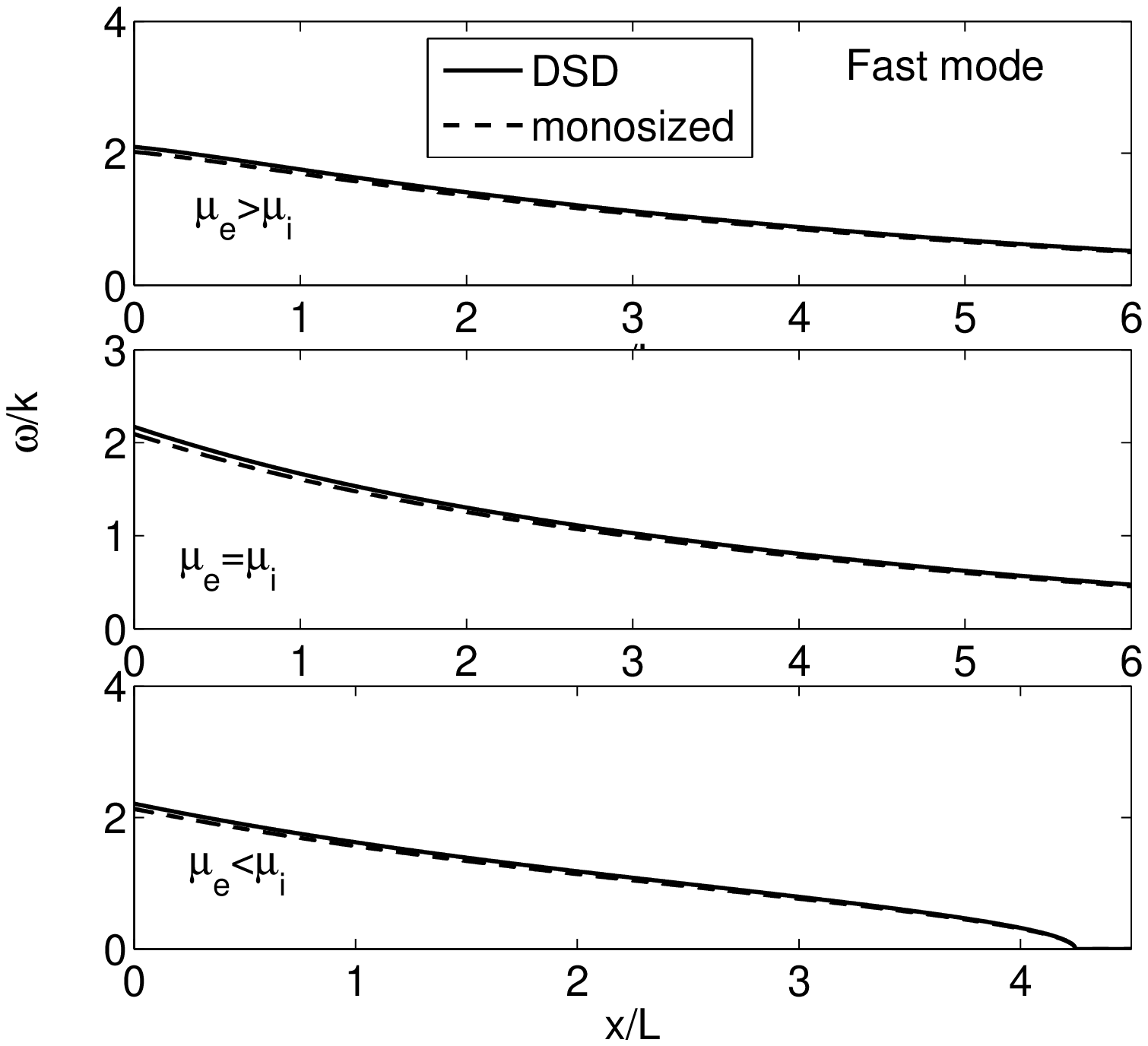}
(b) \includegraphics[height=10cm,width=8cm]{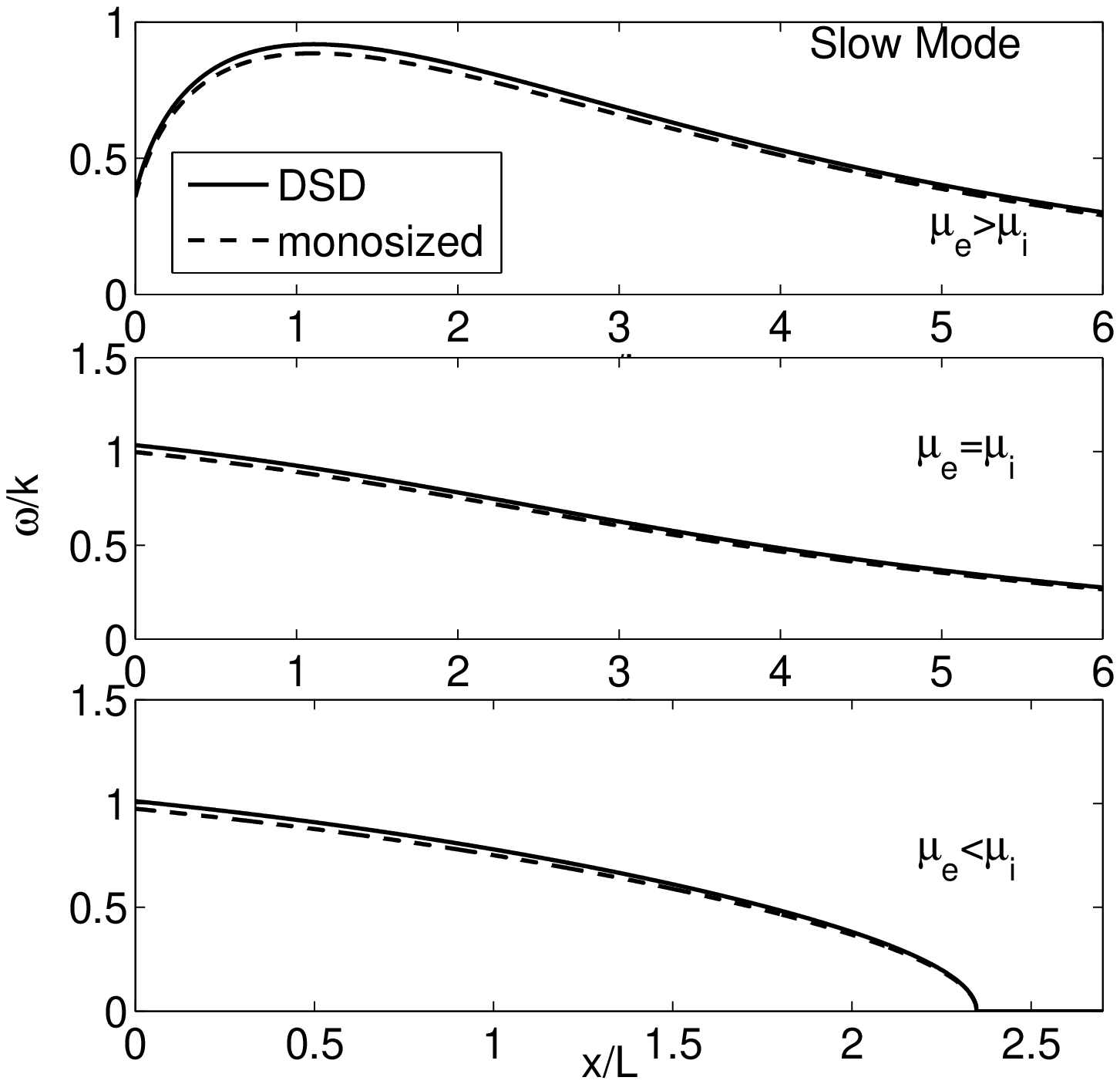}
\caption{ $\omega/k$ vs $x/L$ is plotted for three different cases: $\mu_e=2$, $\mu_i=0.6$ and $\mu_e=0.6$, $\mu_i=0.6$ and $\mu_e=0.01$, $\mu_i=0.6$ in (a) fast mode and (b) slow mode. Here, $k=0.5$ and all other parameters are kept same as in FIG.\ref{Fig: wk_k}}
\label{Fig: wk_xL}
\end{figure*}

\begin{figure*} 
\includegraphics[height=10cm,width=18cm]{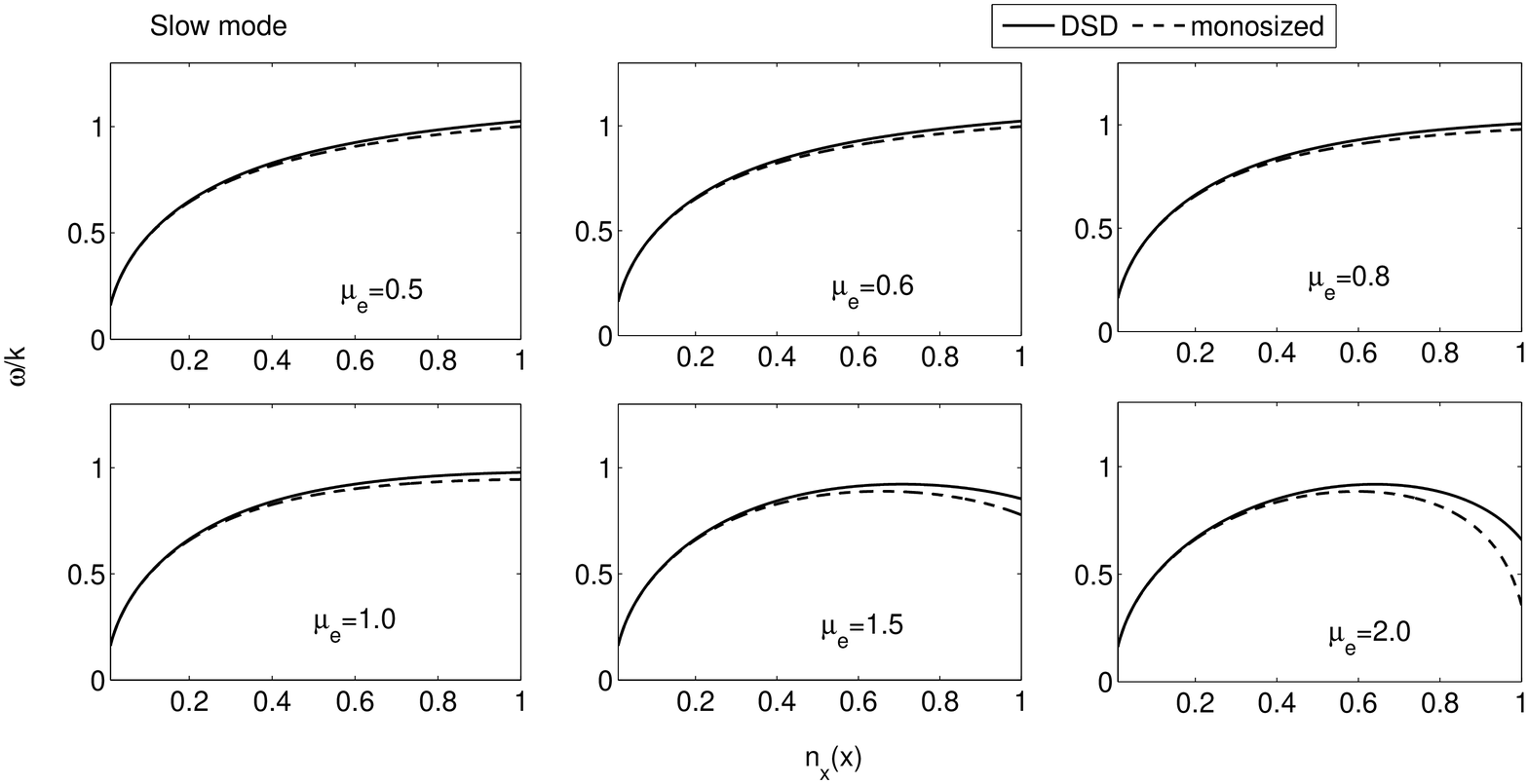}
\caption{ $\omega/k$ vs $n_x(x)$ is plotted for different values of $\mu_e$ slow mode. Here, $k=0.5$, $\mu_i=0.6$ and all other parameters are kept same as in FIG.\ref{Fig: wk_k}}
\label{Fig: wk_nx_slow}
\end{figure*}

\begin{figure*} 
\includegraphics[height=10cm,width=18cm]{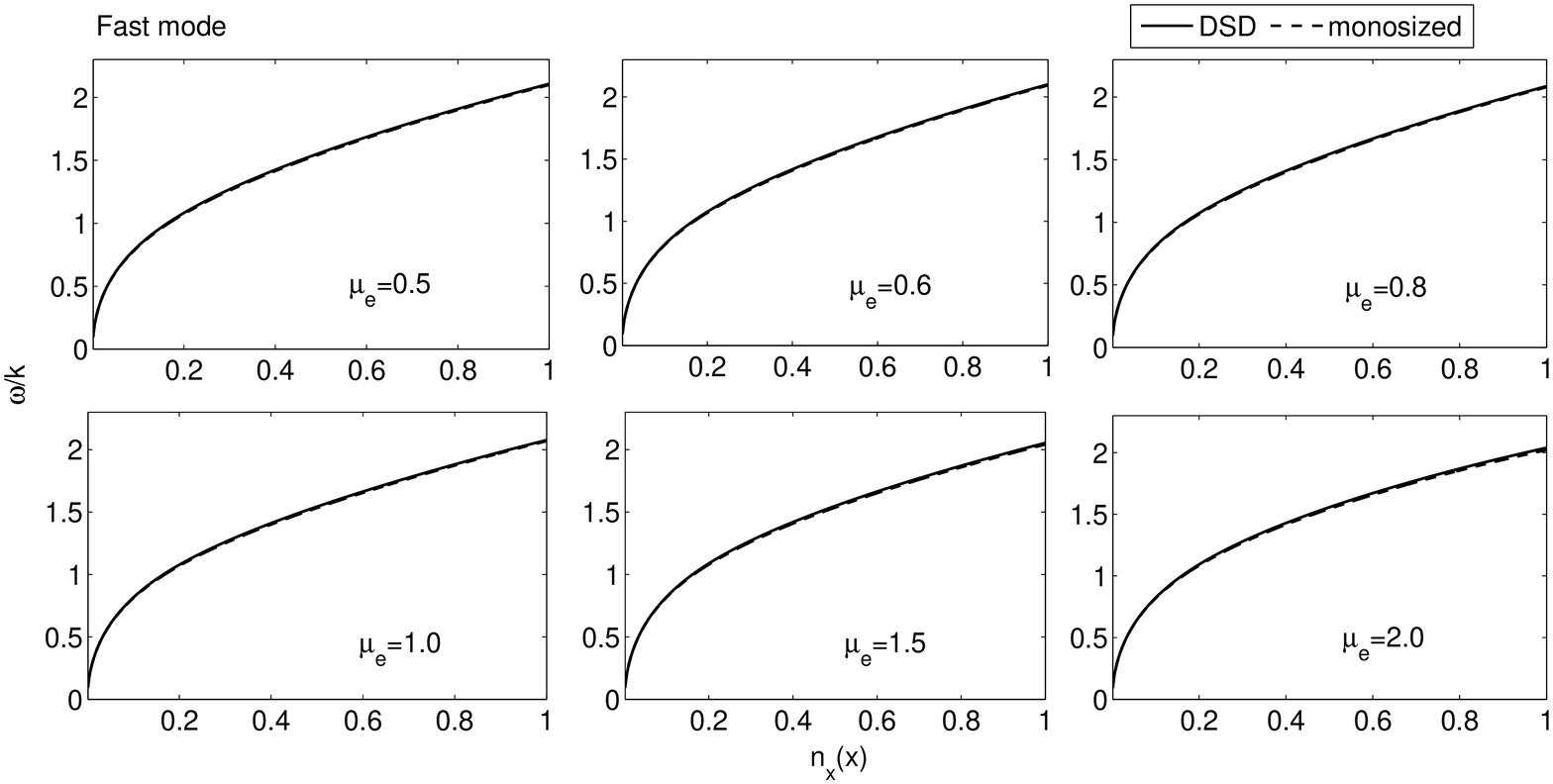}
\caption{ $\omega/k$ vs $n_x(x)$ is plotted for different values of $\mu_e$ fast mode. Here, $k=0.5$, $\mu_i=0.6$ and all other parameters are kept same as in FIG.\ref{Fig: wk_k}}
\label{Fig: wk_nx_fast}
\end{figure*}

\begin{widetext} 
  \begin{eqnarray} \label{dispersion}
  \frac{\omega^2}{k^2} = \frac{1}{4k^2(k^2+\gamma)}\left[(4 a_1 k^2+ a^{\prime \prime}_1+2 a_1 \gamma)\pm \sqrt{(4 a_1 k^2+ a^{\prime \prime}_1+2 a_1 \gamma)^2 -4(k^2+\gamma)(4 a^{2}_1 k^2+ 2 a_1 a^{\prime \prime}_1- a^{\prime 2}_1)} \right]
  \end{eqnarray}
  \end{widetext} 
Instead of DSD if we assume a dusty plasma having monosized dust grain with radius $\overline{a}$, then the phase velocity is given by Eq. (\ref{dispersion}) where $a_1(x)=\overline{a}^{2l-3} N_{tot}(x)$ and $N_{tot}(x)$ is obtained from the charge neutrality condition (\ref{eq3}) as  $N_{tot}(x)=(N_{i0}-N_{e0})\overline{a}^{l}$.
 It is assumed that the equilibrium ions and electrons number densities are obeying exponential growth of the form\cite{taibany2007nonuniform_quantum}
\begin{equation} \label{eq30} 
N_{i0,e0}\left(x\right)=N_{i0,e0}\left(0\right)\mathrm{exp}\mathrm{}(- \mu_{i,e}x/L).
\end{equation} 
 $L$ represents the density scale length and the parameters $\mu_i$, $\mu_e$ are the density gradient scale lengths\cite{taibany2013} for ions and electrons, which may be positive or negative values according to damping or growing number densities, respectively. Here we confined our work for positive values of $\mu_i$, $\mu_e$. For the numerical study of DA waves we have consider the typical dusty plasma parameters from the mesospheric plasmas as follows\cite{mowafy2008, zadorozhny2001}: $n_{e0}\sim 10^3$ $cm^{-3},$ $n_{d0} \sim  10^3$ $cm^{-3},  r \sim 13-40\>nm,$ $Z_d\sim5$, $T_i\sim0.01$ eV, $T_e\sim1$ eV. Equation (\ref{dispersion}) indicates that there are two types of wave mode possible: slow mode corresponding to the negative sign and fast mode corresponding to the positive sign. Let, in Eq. (\ref{dispersion})
 \begin{eqnarray}
 P=4 a_1 k^2+ a^{\prime \prime}_1+2 a_1 \gamma \\
 Q=k^2+\gamma \\
 R=4 a^{2}_1 k^2+ 2 a_1 a^{\prime \prime}_1- a^{\prime 2}_1.
\end{eqnarray}  In both, slow and fast mode, for existence of real values of $\omega/k$, we must have $P^2-4 Q R\geq 0$.  It is clear that  $P^2-4 Q R < 0$ implies $k^2 < -\left[ \gamma+(a^{2}_1-2 a_1 \gamma )^2 /4 a^{2}_1 \right]$ which is not possible as  $\gamma > 0$. Thus  $P^2-4 Q R\geq 0$, $i.e.$, either $P\geq 2\sqrt{Q R}$ or $P\leq -2 \sqrt{Q R}$ under the restriction $R\geq 0$ as $Q\geq 0$. In fast mode, for existence of real value of $\omega/k$, we have  $P+ \sqrt {P^2- 4 Q R}\geq 0$, otherwise the condition $P+ \sqrt{P^2-4 Q R}< 0$ leads to $Q<0$ which is a contradiction.  On the other hand, in slow mode, the condition $P- \sqrt{P^2-4 Q R}\geq 0$ leads to $k^2 \geq \frac{1}{4 a^{2}_1} \left(a^{\prime 2}_1-2 a_1 a^{\prime \prime} \right)$.  We found, there exist a critical wave number, $k_{crit}=\sqrt{\frac{1}{4 a^{2}_1} \left(a^{\prime 2}_1-2 a_1 a^{\prime \prime} \right)}$, below which the phase velocity leads to complex value.  If we consider the the case of homogeneous plasmas, then putting $\mu_e=\mu_i=0$ in Eq. (\ref{eq30}), Eq. (\ref{dispersion}) reduces to 
\begin{equation}
\frac{\omega^2}{k^2}=\frac{(N_{i0}-N_{e0})(l-p+1)(a^{2l-p-2}-1)a^{l-3}_{min}}{(k^2+\gamma)(2l-p-2)(a^{l-p+1}-1)},
\end{equation}
for slow modes, which is same as obtained in our earlier paper on homogeneous plasmas\cite{banerjee2015}. In the case of fast mode the dispersion relation becomes 
\begin{equation}
\frac{\omega^2}{k^2}=\frac{(N_{i0}-N_{e0})(l-p+1)(a^{2l-p-2}-1)a^{l-3}_{min}}{k^2(2l-p-2)(a^{l-p+1}-1)}.
\end{equation}
 In both the cases, for slow and fast modes, it is clear from Eq. (\ref{eq30}) that, as the wave number $k\rightarrow0$ the phase velocity $\omega/k \rightarrow\infty$  and when $k \rightarrow\infty$, $\omega/k\rightarrow0$. The variation of the phase velocity $\omega/k$ has been plotted with the increasing wave number $k$ for different values of $x/L$ in Fig.\ref{Fig: wk_k}. It shows that initially phase velocity decreases rapidly and then decreases gradually towards zero.
 The phase velocities are plotted against $x/L$ in Fig.\ref{Fig: wk_xL} for both the cases of slow and fast mode. Here, the doted lines and solid lines represent the curves for monosized and multisized dust grains, respectively. The phase velocities are plotted for the three different cases $:$ $\mu_e>\mu_i$, $\mu_e=\mu_i$ and $\mu_e<\mu_i$. In all these three cases, both in slow and fast mode, it has been noticed that due to DSD the phase velocity increases in comparison to the case of monosized dust. It is found that in fast mode, for $\mu_e \geq \mu_i$, the phase velocity decreases along $x/L$ but in slow mode, for $\mu_e>\mu_i$, initially the phase velocity increases and after crossing a critical value in terms of $x/L$ it stars decreasing gradually. However, in slow modes, as $\mu_e \rightarrow \mu_i$,  this initial increment in phase velocities has not been observed. For $\mu_e<\mu_i$, both in slow and fast mode, it has been found that above a critical value of $x/L$ the phase velocity becomes imaginary and so wave damping takes place. The obtained results for slow mode have been found similar to our earlier work on nonlinear waves in inhomogeneous dusty plasmas\cite{banerjee2017}.
The curves in Fig.\ref{Fig: wk_nx_slow} and Fig.\ref{Fig: wk_nx_fast} represents the phase velocities  for different values of $\mu_e$ in both the modes: slow and fast, respectively. The dotted and solid lines are showing the plots of phase velocities for monosized and multisized dust grains, respectively. In slow mode, as $\mu_e$ increases small increments in phase velocities have been noticed  along increasing dust densities in comparison to the case of monosized dust grains. But for the case of fast mode these increments are negligibly small. It has been observed that the phase velocities increase along the increasing dust density for  $\mu_e \leq 1$, but for $\mu_e>1$, initially the phase velocity increases and after crossing a critical dust number density it starts decreasing, whereas in fast mode the phase velocity increases always along increasing dust density. 

\section{CONCLUSIONS}
In this work, the linear analysis of dust acoustic waves have been carried out in an inhomogeneous dusty plasma having multisized negatively charged dust grains and Boltzman electrons and ions. Here, the dust grains are considered to  follow power law dust size distribution. The spatial inhomogeneity in equilibrium density profiles have been considered. Assuming the spatial and time dependent perturbed quantities varies as $A(x,t)=A_1(x) e^{-i \omega t}$, a dispersion relation suitable for inhomogeneous  plasmas has been derived. Two different modes of wave propagation: slow and fast, have been discussed. In the case of slow mode, a critical point in terms of wave number has been derived below which wave damping occurs. Also, due to number density inhomogeneity, when the density gradient scale length of ions is large in comparison to electrons, the wave damping occurs after crossing a critical spatial distance.

\section*{Acknowledgements}
One of the authors acknowledge the financially support by UGC New Delhi under the Dr. D.S. Kothari Post Doctoral Fellowship Scheme.

\include{ref}
\bibliography{Manuscript}

\end{document}